\documentclass{aastex701}

\usepackage{graphicx}
\usepackage{color}
\usepackage{dcolumn}
\usepackage{bm}
\usepackage{hyperref}
\usepackage{subfigure}
\usepackage{acronym}
\usepackage{CJK}
\usepackage{orcidlink}
\usepackage{braket}
\usepackage{amsmath}

\graphicspath{{./}{figures/}}

\newacro{GR}{general relativity}
\newacro{GW}{gravitational wave}
\newacro{BH}{black hole}
\newacro{BBH}{binary black hole}
\newacro{QNM}{quasi-normal mode}
\newacro{QQNM}{quadratic \ac{QNM}}
\newacro{SNR}{signal-to-noise ratio}
\newacro{IMR}{inspiral-merger-ringdown}
\newacro{NR}{numerical relativity}
\newacro{KLD}{Kullback-Leibler divergence}
\newacro{LVK}{LIGO-Virgo-KAGRA}

\begin{document}
\begin{CJK*}{UTF8}{gbsn}

\title{Constraining the Quadratic-mode Amplitude Coupling in GW250114}

\author[orcid=0009-0008-9388-0620]{Yuxin Yang(杨雨鑫)}
\affiliation{MOE Key Laboratory of TianQin Mission \\ TianQin Research Center for Gravitational Physics \& School of Physics and Astronomy \\ Frontiers Science Center for TianQin Gravitational Wave Research Center of CNSA \\ Sun Yat-sen University (Zhuhai Campus), Zhuhai 519082, China}
\email{yangyx97@mail2.sysu.edu.cn}

\author[0000-0002-2252-3131]{Changfu Shi(石常富)} 
\affiliation{MOE Key Laboratory of TianQin Mission \\ TianQin Research Center for Gravitational Physics \& School of Physics and Astronomy \\ Frontiers Science Center for TianQin Gravitational Wave Research Center of CNSA \\ Sun Yat-sen University (Zhuhai Campus), Zhuhai 519082, China}
\email[show]{shichf6@mail.sysu.edu.cn}

\author[0000-0002-7869-0174]{Yi-Ming Hu(胡一鸣)}
\affiliation{MOE Key Laboratory of TianQin Mission \\ TianQin Research Center for Gravitational Physics \& School of Physics and Astronomy \\ Frontiers Science Center for TianQin Gravitational Wave Research Center of CNSA \\ Sun Yat-sen University (Zhuhai Campus), Zhuhai 519082, China}
\email[show]{huyiming@sysu.edu.cn}

\begin{abstract}

    Detecting quadratic quasi-normal modes in black hole ringdowns would provide evidence for nonlinear gravitational dynamics, while measuring their properties would enable tests of the corresponding predictions of general relativity. 
    Specifically, second-order black hole perturbation theory predicts that their amplitudes scale with the product of the amplitudes of their parent linear modes, with a coupling coefficient that depends on the spin of the remnant black hole.
    This mode-specific coupling coefficient has not yet been directly measured from gravitational wave data.
    Here, we use Bayesian inference on the GW250114 ringdown, modeled with the $220$, $221$, and $220\times220$ modes, to infer the amplitude-coupling coefficient of the $220\times220$ mode.
    The inferred coefficient is consistent with predictions from numerical relativity fits and second-order perturbation theory; no comparison shows a deviation exceeding $1.3 \sigma$.
    Although the current data cannot distinguish among these spin-dependent predictions, this measurement establishes a basis for more stringent observational tests of nonlinear gravitational dynamics in black-hole ringdowns.

\end{abstract}



\section{Introduction}

Since the detection of GW150914~\citep{LIGOScientific:2016aoc}, hundreds of \ac{GW} events from \ac{BBH} have been observed~\citep{LIGOScientific:2018mvr, LIGOScientific:2020ibl, KAGRA:2021vkt, LIGOScientific:2025slb}.
Analyzing the ringdown signal in these events has become an important way to test gravitational theories in the strong field regime.
The dominant part of the ringdown signal can be described as a superposition of \acp{QNM}~\citep{Nollert:1999ji,Kokkotas:1999bd,Berti:2009kk,Berti:2005ys,Konoplya:2011qq}.
For a Kerr \ac{BH} in \ac{GR}, the mass and spin determine its complex frequencies.
Measuring these frequencies and comparing them with values predicted from the remnant mass and spin therefore tests the \ac{BH} no-hair theorem and, more broadly, \ac{GR}~\citep{Dreyer2004,Gossan:2011ha,Shi:2019hqa,LIGOScientific:2026wpt}.
In addition to \acp{QNM}, the ringdown signal may also contain nonlinear contributions, for example, \acp{QQNM}~\citep{Okuzumi:2008ej, Nakano:2007cj, London:2014cma, Mitman:2022qdl, Cheung:2022rbm, Baibhav:2023clw}.
Detecting \acp{QQNM} would provide evidence for nonlinear gravitational effects in the strong-field regime, while measuring their properties could deepen our understanding of \ac{BH} perturbations.
As \acp{QQNM} arise from the coupling of linear \acp{QNM}, they also offer a way to study how gravitational waves can themselves source further gravitational radiation through nonlinear interactions.
Forecasting studies have also assessed the detectability of \acp{QQNM} with next-generation detectors and their potential contribution to black-hole spectroscopy~\citep{Yi:2024elj,Lagos:2024ekd}.

Analyses of \ac{BH} ringdown signals are beginning to extend beyond linear mode identification to searches for \acp{QQNM}.
Progress beyond the dominant mode initially came from searches for additional linear \acp{QNM}, with suggestive but inconclusive evidence reported for high-\ac{SNR} events such as GW150914~\citep{Isi:2019aib,Cotesta:2022pci, Isi:2022mhy, Wang:2024yhb, Wang:2023ljx}, GW190521~\citep{Capano:2021etf, LIGOScientific:2020iuh, Siegel:2023lxl}, GW231028~\citep{Wang:2025baj}, GW231123~\citep{Wang:2025rvn}, and GW230814~\citep{Tang:2025jyj}.
The exceptionally high \ac{SNR} of GW250114 enabled a significant advance: the \ac{LVK} Collaboration reported decisive evidence for the $221$ overtone~\citep{LIGOScientific:2025rid,LIGOScientific:2025wao}.
The collaboration also identified a contribution from the $m=\lvert\ell\rvert=4$ multipole~\citep{LIGOScientific:2025wao}.
Subsequent studies investigated whether this multipolar contribution could originate from \acp{QQNM}.
In a ringdown-only analysis, \citet{Yang:2025ror} found a conditional preference of  Bayes factor $\mathcal{B}\simeq10$ for $220+221+220Q$ over $220+221+440$, although $220+221$ remained favored overall.
Using \ac{IMR} priors and a bundled model of ten \acp{QQNM}, \citet{Wang:2026rev} obtained $\mathcal{B}=62$ in favor of the quadratic contribution.
Other potential contributions to the early post-merger signal have also been explored.
\citet{LIGOScientific:2025wao} searched for the second overtone, $222$, but found no conclusive evidence for its presence.
Beyond \acp{QNM}, \citet{Lu:2025vol} reported observational evidence for a direct wave component in GW250114.

Recent searches have used a prediction of second-order \ac{BH} perturbation theory to identify candidate \acp{QQNM}: the complex frequency of a quadratic mode equals the sum of those of its parent linear modes~\citep{Nakano:2007cj,Besson:2024adi,Yang:2025ror,Wang:2026rev}.
This frequency-sum relation, however, is not unique to \ac{GR}: it also arises in a simplified nonlinear wave model and is expected to persist in perturbative modified-gravity frameworks~\citep{Okuzumi:2008ej,Silva:2024ffz}.
Measuring the amplitude coupling coefficient between a \ac{QQNM} and its parent linear modes therefore can provide a more discriminating test of \ac{GR}.

Within \ac{GR}, the amplitude coupling of \acp{QQNM} has been studied using both \ac{NR} simulations and second-order \ac{BH} perturbation theory.
\cite{Mitman:2022qdl} and \cite{Cheung:2022rbm} independently identified \acp{QQNM} for the first time in \ac{NR} simulations, and estimated the coupling coefficient relating the amplitude of a quadratic mode to the product of the amplitudes of its parent linear modes.
Building on these results, \cite{Cheung:2023vki} analyzed a larger set of \ac{NR} simulations and found that this ratio depends on the remnant \ac{BH} spin, for which they provided a linear fitting formula.
\cite{Redondo-Yuste:2023seq} further proposed a nonlinear fitting formula that gives a better fit in the high-spin regime, and showed that the ratio is largely independent of the initial conditions of the \ac{BBH} merger.
In parallel, several studies have attempted to compute the spin dependence of the amplitude ratio directly from second-order perturbation theory, rather than by fitting \ac{NR} simulations~\citep{Bucciotti:2024zyp, Ma:2024qcv, Zhu:2024rej, Bucciotti:2024jrv, Bourg:2024jme, Mitman:2025hgy}.
Following these exploratory studies, \citet{Khera:2024bjs} derived a spin-dependent relation that agrees with the \ac{NR} simulations.
Together, these studies provide predictions for the remnant-spin dependence of the \ac{QQNM} amplitude ratio within \ac{GR}, which can serve as benchmarks for observational tests of nonlinear predictions of \ac{GR}.

Here, we extend observational studies of \acp{QQNM} from mode identification to an observational consistency test of amplitude coupling.
We perform Bayesian inference on the GW250114 ringdown to measure the amplitude-coupling coefficient of the $220\times220$ mode, which is expected to be the dominant \ac{QQNM} in \ac{BBH} ringdowns~\citep{Mitman:2022qdl,Cheung:2022rbm}. 
Comparing the inferred posterior with representative predictions~\citep{Cheung:2022rbm,Cheung:2023vki,Redondo-Yuste:2023seq,Khera:2024bjs}, we find no statistically significant discrepancy within the current uncertainties.

The remainder of this work is organized as follows. 
Section~\ref{sec:ringdown_model} introduces the ringdown model and the definition of the nonlinear amplitude ratio. 
Section~\ref{sec:setup} describes the Bayesian inference setup for GW250114. 
Section~\ref{sec:result} presents the inferred amplitude ratio and its comparison with theoretical predictions, and Section~\ref{sec:discussion} summarizes our conclusions.

\section{Ringdown model}\label{sec:ringdown_model}

According to \ac{BH} perturbation theory, the ringdown phase signal following a \ac{BBH} merger can be described as \acp{QNM}. Indexing by the angular-mode numbers $\ell$ and $m$, and the radial overtone number $n$, the linear \acp{QNM} can be written as \citep{Berti:2009kk}
\begin{equation}
    h_{+} - \mathrm{i} h_{\times} = \frac{M_z}{D_L} \sum_{lmn} {}_{-2} S_{\ell mn} \tilde{A}_{\ell mn} \mathrm{e}^{\mathrm{i} \tilde{\omega}_{\ell mn} (t-t_0)},
    \label{eq:ringdown_model}
\end{equation}
where $M_z$ and $D_L$ represent the redshifted mass of the remnant black hole and the luminosity distance of the system, respectively.
$\tilde{A}_{\ell mn} = A_{\ell mn} \mathrm{e}^{\mathrm{i}\phi_{lmn}}$ and $\tilde{\omega}_{\ell mn}=\omega_{\ell mn}+\mathrm{i} / \tau_{\ell mn}$ denote the complex amplitudes and frequencies of each mode.
${}_{-2} S_{lmn}(\iota, \varphi; \chi_f)$ are the -2 weighted spin spheroidal harmonics, which depend on the inclination $\iota$, the azimuthal angle $\varphi$ and the remnant spin $\chi_f$~\citep{Berti:2005gp}.
Assuming a Kerr \ac{BH}, the ringdown frequencies can be obtained by solving the Teukolsky equation \citep{Teukolsky:1973ha}, and their values depend on the mass $M_z$ and dimensionless spin $\chi_f$ of the remnant \ac{BH}.
By contrast, the amplitudes depend on the dynamical evolution of the progenitors.

Second-order \ac{BH} perturbation theory predicts the existence of \acp{QQNM}, which arise from the nonlinear coupling of linear modes.
The complex frequencies of these quadratic modes are determined by those of corresponding linear modes and can be written in the form~\citep{Mitman:2022qdl, Cheung:2022rbm, Nakano:2007cj}
\begin{equation}
    \begin{aligned}
        \omega_{\ell_{1}m_{1}n_{1} \times \ell_{2}m_{2}n_{2}} &= \omega_{\ell_{1}m_{1}n_{1}} + \omega_{\ell_{2}m_{2}n_{2}}, \\
        \tau_{\ell_{1}m_{1}n_{1} \times \ell_{2}m_{2}n_{2}}^{-1} &= \tau_{\ell_{1}m_{1}n_{1}}^{-1} + \tau_{\ell_{2}m_{2}n_{2}}^{-1}, \\
    \end{aligned}
    \label{eq:nonlinear_frequency}
\end{equation}
where $\ell_{1}m_{1}n_{1}$ and $\ell_{2}m_{2}n_{2}$ denote the pair of linear modes that generate the quadratic mode.
For the $220\times220$ contribution considered in this work, we model its angular dependence using the $(\ell,m,n)=(4,4,0)$ Kerr spheroidal harmonic, ${}_{-2}S_{440}(\iota,\varphi;\chi_f)$.

The complex amplitude of a \ac{QQNM} is proportional to the product of the complex amplitudes of the corresponding linear modes.
We characterize this relation through the complex amplitude ratio
\begin{equation}
    \tilde{\mathcal{R}}_{\ell_{1}m_{1}n_{1}\times \ell_{2}m_{2}n_{2}} = \frac{\tilde{A}_{\ell_{1}m_{1}n_{1}\times \ell_{2}m_{2}n_{2}}}{\tilde{A}_{\ell_{1}m_{1}n_{1}} \tilde{A}_{\ell_{2}m_{2}n_{2}}}.
    \label{eq:nonlinear_ratio}
\end{equation}
Because the complex frequency of the quadratic mode is the sum of those of its parent modes, the numerator and denominator of Eq.~\ref{eq:nonlinear_ratio} acquire identical phase and damping factors under a shift of the reference time.
The ratio $\tilde{\mathcal{R}}$ is therefore independent of the reference time.
Consequently, analyses performed with different ringdown start times probe the same underlying coupling ratio, although their inferred posteriors may differ because of statistical uncertainties and modeling systematics.

Current numerical relativity analyses~\citep{Cheung:2023vki, Redondo-Yuste:2023seq} and theoretical studies~\citep{Bucciotti:2024jrv, Bucciotti:2024zyp, Zhu:2024rej, Bourg:2024jme, Khera:2024bjs} both indicate that the ratio should be determined by the remnant \ac{BH} 's dimensionless spin $\chi$.

\section{GW250114 Analysis Setup}\label{sec:setup}

We perform a time-domain analysis~\citep{Isi:2019aib,Isi:2021iql} of the post-peak signal of GW250114, using publicly available strain data from the Gravitational Wave Open Science Center~\citep{LIGOScientific:2026jgl}.
The measured GPS peak arrival times of GW250114 are $t_{\rm peak}^{\rm LHO}=1420878141.2190^{+0.0001}_{-0.0001}\,\mathrm{s}$ at LIGO Hanford and
$t_{\rm peak}^{\rm LLO}=1420878141.2165^{+0.0001}_{-0.0001}\,\mathrm{s}$ at LIGO Livingston.
For our analysis, we adopt the geocentric reference peak time $t_{\rm peak}=1420878141.235932\,\mathrm{s}$~\citep{LIGOScientific:2025rid}.
A truncated QNM model is expected to describe the signal adequately only sufficiently late after merger, but the onset of its applicability remains uncertain~\citep{Nollert:1996rf, Bhagwat:2017tkm, Clarke:2024lwi, Baibhav:2023clw}.
We therefore repeat the analysis at different start times, defined by $\Delta t=t_0-t_{\rm peak}$.
These start times are spaced by $t_{M_z}=0.337$ ms, as determined from the redshifted mass $M_z$ of the remnant \ac{BH}.
To set the ringdown start time consistently in each detector, we adopt the reference sky location associated with the maximum-likelihood \ac{IMR} waveform analysis of GW250114, fixing $\alpha = 2.333$ rad and $\delta=0.190$ rad~\citep{LIGOScientific:2025rid,LIGOScientific:2025wao}.
We use the corresponding reference polarization angle $\psi = 1.329$~\citep{LIGOScientific:2025rid}.
Because the luminosity distance and inclination are strongly degenerate with the mode amplitudes, we also fix them to $D_L=411$ Mpc and $\iota=0.76$ rad~\citep{LIGOScientific:2026ctl}.
We use \texttt{bilby}~\citep{bilby_paper, Bilby_17371955, Romero-Shaw:2020owr} and \texttt{dynesty}~\citep{Speagle:2019ivv, sergey_koposov_2025_17268284} to perform our Bayesian inference with 1000 live points to sample the posterior distribution and stop when the change in the logarithm of the Bayesian evidence is less than $0.1$.

In this work, we analyze the data using a ringdown model that includes only three \acp{QNM}: $220$, $221$, and $220\times 220$.
The linear modes are described by Eq.~\ref{eq:ringdown_model}, while the nonlinear mode is described by Eqs.~\ref{eq:nonlinear_frequency} and~\ref{eq:nonlinear_ratio}.
All complex frequency parameters are determined by the mass $M_z$ and spin $\chi_f$ of the remnant \ac{BH}, and each complex amplitude is decomposed into two free parameters: an absolute value and a phase.
The final model therefore contains eight parameters: $M_z$, $\chi_f$, $A_{220}$, $\phi_{220}$, $A_{221}$, $\phi_{221}$, $\mathcal{R}_{220\times 220}$, and $\phi_{220 \times 220}$.
The phase $\phi_{220 \times 220}$ depends on the definition of the merger time and is therefore difficult to determine robustly.
We therefore focus on measuring $\mathcal{R}_{220\times 220}$, denoted by $\mathcal{R}$ in the following for convenience.

\begin{figure*}
    \centering
    \includegraphics[width=0.8\linewidth]{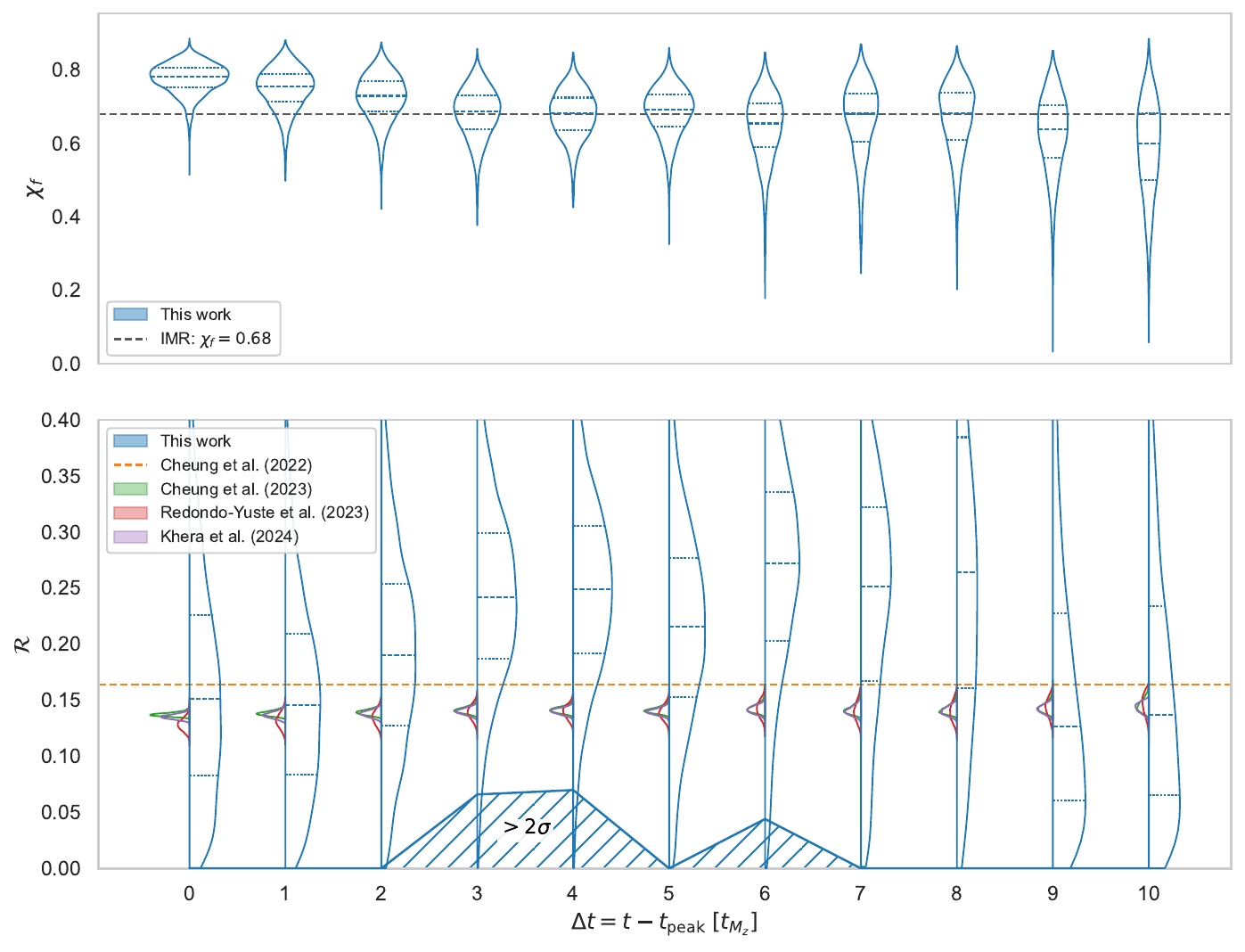}
    \caption{Time dependence of the inferred spin of the remnant \ac{BH} and the quadratic-mode amplitude ratio for GW250114. The upper panel shows the posterior distribution of the remnant \ac{BH} dimensionless spin $\chi_f$ vs. the analysis start time $\Delta t$. The gray dashed line marks the reference maximum-likelihood value from the \ac{IMR} waveform. The lower panel shows the corresponding amplitude ratio $\mathcal{R}$ for the $220 \times 220$ mode. At each start time, the right half of the blue violin gives the observed posterior of $\mathcal{R}$, while the left half of the violin shows the theoretical predictions obtained by mapping the same $\chi_f$ posterior through the spin-dependent relations of \cite{Cheung:2023vki}, \cite{Redondo-Yuste:2023seq}, and \cite{Khera:2024bjs}. The orange dashed line denotes the spin-independent prediction of \cite{Cheung:2022rbm}. Internal dashed lines mark the quartiles of the plotted posterior distributions. The hatched region in the lower panel is excluded at least $2\sigma$; regions excluded at higher credible levels lie within it. The broad observed distributions and their start-time dependence indicate that the current measurement is consistent with the overall theoretical scale but does not distinguish decisively among the different amplitude-ratio models.}
    \label{fig:ratio_chi_vs_start_time}
\end{figure*}

\begin{figure*}
    \centering
    \includegraphics[width=0.8\linewidth]{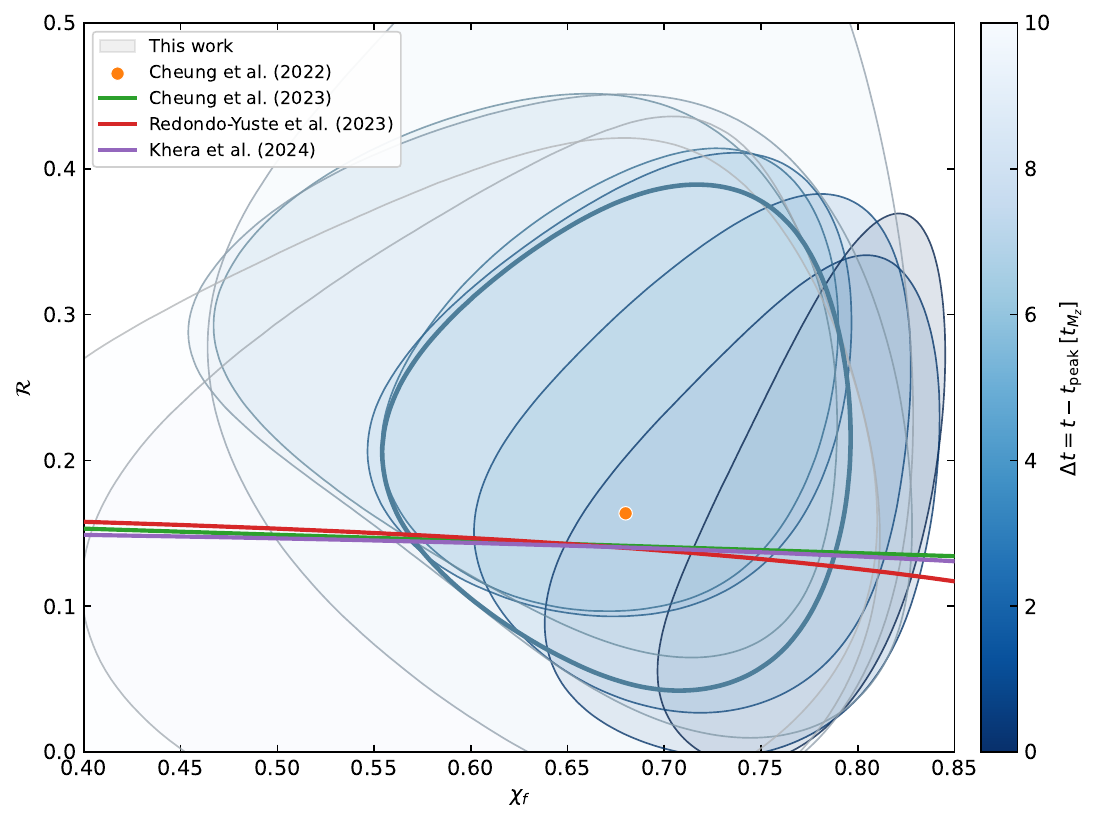}
    \caption{Posterior constraints on the quadratic-mode amplitude ratio in the ($\chi_f$, $\mathcal{R}$) plane. The translucent blue areas show the $80\%$ two-dimensional credible regions of the GW250114 posteriors for different analyses start times, with darker shades indicating earlier times and a thick contour highlighting $\Delta t=5\,t_{M_z}$. The earlier start times are drawn with the darker contours and later start times with lighter contours. The orange point denotes the spin-independent prediction of \cite{Cheung:2022rbm}, evaluated at $\chi_f = 0.68$. The green, red, and purple curves show the spin-dependent predictions of ~\cite{Cheung:2023vki}, ~\cite{Redondo-Yuste:2023seq}, and ~\cite{Khera:2024bjs}, respectively. The broad posterior support indicates that the inferred amplitude ratio of the $220\times 220$ mode is consistent with the theoretical scale, but the current data do not decisively distinguish among the different spin-dependent models. A three-dimensional visualization of these results, with interactive controls available online, is provided in Appendix.}
    \label{fig:ratio_chi_theory_overlay}
\end{figure*}

\section{Results}\label{sec:result}

We compare the inferred amplitude ratio with several theoretical predictions for the $220\times220$ quadratic mode. 
The early study identified nonlinear modes in \ac{NR} simulations of equal-mass, nonspinning \ac{BBH} mergers and provided a spin-independent constant estimate for $\mathcal{R}$~\citep{Cheung:2022rbm}. 
More recent studies suggest that the coupling strength depends on the remnant spin. 
In particular, \cite{Cheung:2023vki} proposed a linear spin-dependent relation calibrated to \ac{NR} waveforms, while \cite{Redondo-Yuste:2023seq} found an improved fit with a nonlinear dependence relation. 
Using higher-order \ac{BH} perturbation theory, \cite{Khera:2024bjs} provided a spin-dependent prediction that is close to the \ac{NR} result.
Together, these models provide representative predictions for how the nonlinear amplitude ratio should vary with $\chi_f$ in \ac{GR}.

We repeat the analysis for a sequence of start times after the strain peak and summarize the results in Fig.~\ref{fig:ratio_chi_vs_start_time}.
The upper panel shows the inferred remnant spin and compares it with the reference value obtained from the \ac{IMR} waveform analysis.
When the start time is close to the peak time, the inferred $\chi_f$ shows a bias from the reference line.
This discrepancy suggests that this portion of the data may not be adequately described by a ringdown model containing only the three QNMs considered here.
As the analysis start time moves beyond $3\,t_{M_z}$, the inferred spin tends to become more consistent with the reference value, suggesting that our model begins to provide a more suitable description of the later signal.

The lower panel of Fig.~\ref{fig:ratio_chi_vs_start_time} compares the inferred amplitude ratio $\mathcal{R}$ with the theoretical predictions.
The posterior distributions are generally broad, reflecting the limited \ac{SNR} of the GW250114 post-peak data.
The spin-dependent predictions are obtained by mapping the $\chi_f$ posterior through the corresponding theoretical relations, while the spin-independent estimate of \citet{Cheung:2022rbm} is shown as a horizontal line.
Overall, the observed distributions encompass the theoretical predictions.
To quantify this agreement, we take the median of each theoretical distribution as its representative value, compute its percentile within the observed posterior distribution of $\mathcal{R}$, and convert this percentile to an equivalent Gaussian significance.
Across all analysis start times and theoretical models considered, the largest deviation is approximately $1.3\sigma$, corresponding to a two-sided Gaussian-equivalent tail probability of about $19\%$.
This broad overlap also indicates that the current observation does not decisively favor any individual model.

We also assess the posterior support away from $\mathcal{R}=0$ using a boundary-corrected estimate of the marginal posterior density with \texttt{GetDist}~\citep{Lewis:2019xzd}.
We integrate the density over regions where it exceeds its value at zero and convert the resulting probability to a two-sided Gaussian-equivalent credible level.
The hatched region in Fig.~\ref{fig:ratio_chi_vs_start_time} marks values below the lower boundary of the $2\sigma$ highest posterior density region.
The support away from zero is stronger for $\Delta t=3\sim 6\,t_{M_z}$, reaching a maximum of $2.7\sigma$ at $\Delta t=4\,t_{M_z}$.
For $\Delta t\geq9\,t_{M_z}$, however, zero cannot be excluded even at the $1\sigma$ level.
This loss of constraining power is consistent with the
decay of the signal, making the quadratic-mode amplitude
increasingly difficult to distinguish from noise.

To visualize the joint dependence on the remnant spin and the amplitude ratio, Fig.~\ref{fig:ratio_chi_theory_overlay} shows the $80\%$ credible region posterior in the $(\chi_f,\mathcal{R})$ plane.
The shaded regions are color-coded by analysis start time, with a thicker boundary marking the result at $\Delta t=5\,t_{M_z}$.
We highlight this start time for consistency with the subsequent injection studies, as it lies within the interval of stronger posterior support away from $\mathcal{R}=0$ while reducing potential contamination from early-time contributions, including the direct wave.
The posterior support spans the region covered by the existing theoretical predictions, with a tendency for the inferred amplitude ratio to extend above the central theoretical curves.
However, the broad two-dimensional posterior and the variation among start times indicate that this trend is not statistically decisive.
We therefore interpret the result as being consistent with the current \ac{GR} predictions for the $220\times220$ quadratic-mode amplitude ratio, while not yet providing enough precision to distinguish among the different spin-dependent relations.

\begin{figure*}
    \centering
    \includegraphics[width=0.8\linewidth]{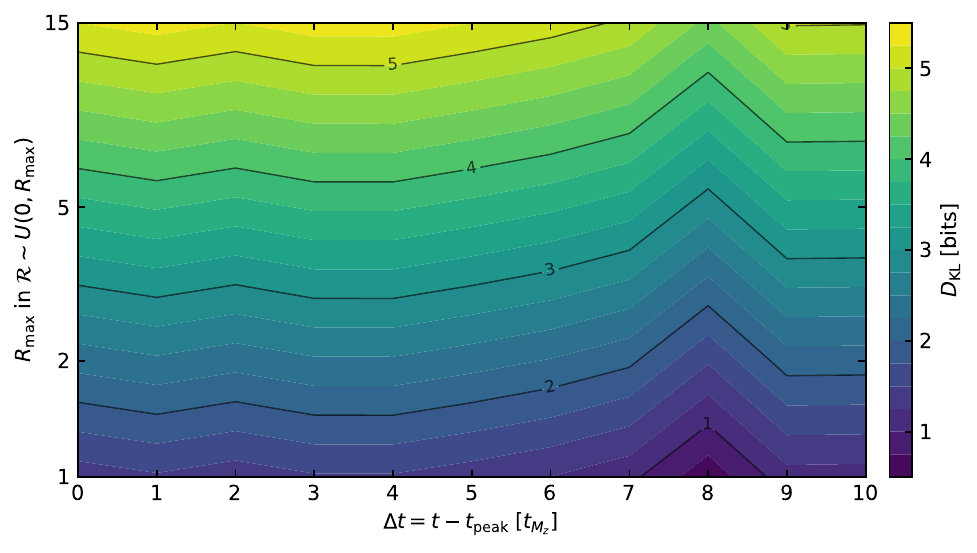}
    \caption{
    Information gain for the inferred amplitude ratio $\mathcal{R}$ as a function of ringdown start time $\Delta t$ and the upper bound $R_{\max}$ of the uniform prior $\mathcal{R} \sim U(0, R_{\max})$. Colors and labeled contours show $D_{\mathrm{KL}}$ in bits. The information gain decreases for narrower priors and shows a dip near $\Delta t=8\,t_{M_z}$.
    }
    \label{fig:ratio_information_gain}
\end{figure*}

\begin{figure*}
    \centering
    \includegraphics[width=0.8\linewidth]{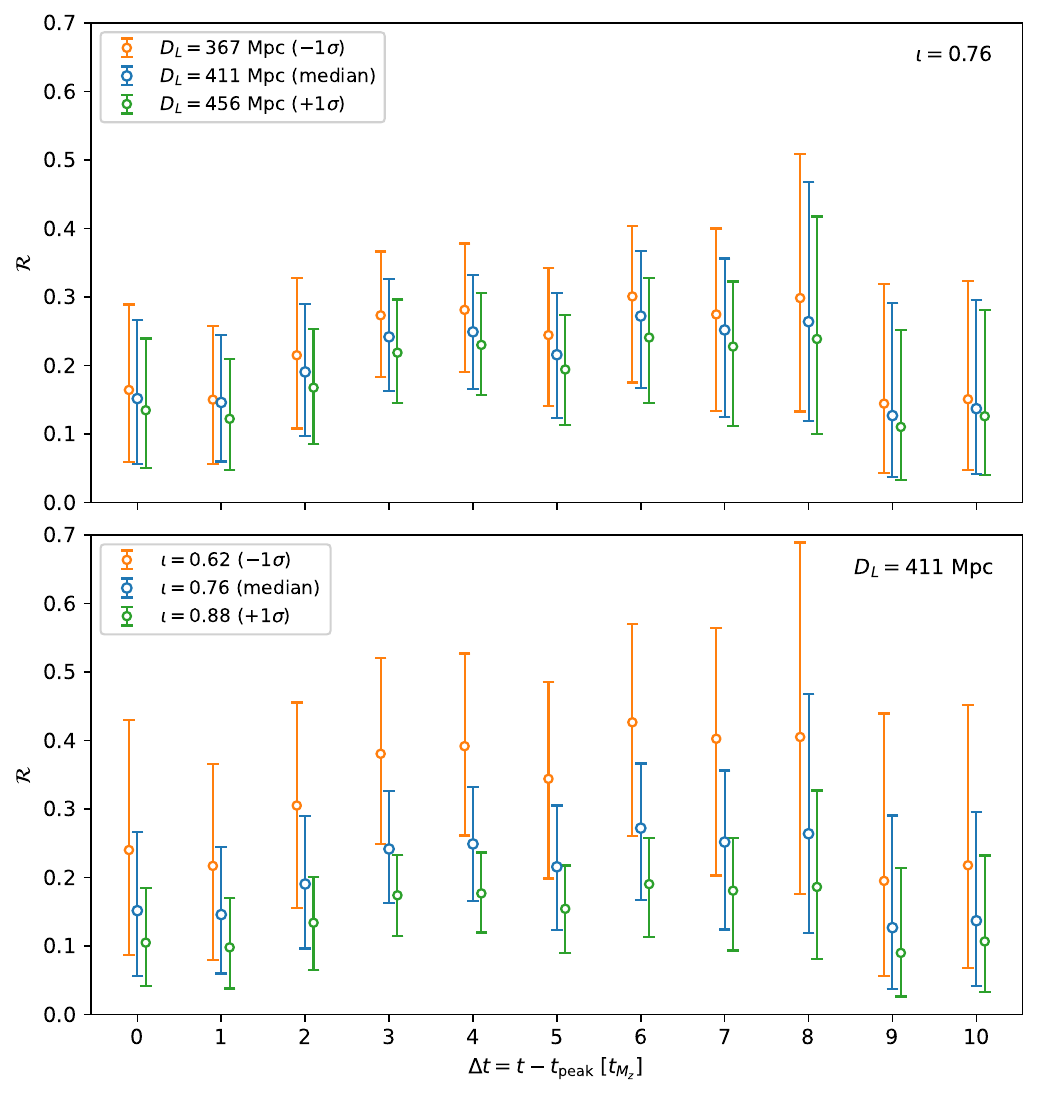}
    \caption{
    Sensitivity of the inferred quadratic-mode amplitude ratio $\mathcal{R}$ to the fixed extrinsic parameters in the GW250114 ringdown analysis.
    The markers show the posterior median of $\mathcal{R}$, and the error bars denote the central $68\%$ credible interval, as a function of the ringdown start time $\Delta t=t-t_{\rm peak}$ in units of $t_{M_z}$.
    The upper panel varies the fixed luminosity distance, $D_L=367$, $411$, and $456\,{\rm Mpc}$, while keeping the inclination fixed at $\iota=0.76$.
    The lower panel varies the fixed inclination, $\iota=0.62$, $0.76$, and $0.88$, while keeping $D_L=411\,{\rm Mpc}$.
    The similar start-time dependence across these choices indicates that the qualitative behavior of the inferred amplitude ratio is stable, while the systematic shifts in normalization show that the measurement remains conditional on the assumed distance and inclination.
    }
    \label{fig:ratio_parameter_sensitivity}
\end{figure*}

Although the inferred posteriors do not yet distinguish among the theoretical predictions, they still constrain $\mathcal{R}$ relative to the prior.
To quantify this constraint and facilitate comparisons with other methods using the same priors, we compute the information gain as the \ac{KLD} between the posterior and several uniform priors on $\mathcal{R}$~\citep{KullbackLeibler1951}.
All values are derived from a single inference performed with the broadest prior, $\mathcal{R}\sim U(0,15)$; for each smaller $R_{\max}$, the posterior samples are reweighted to the corresponding prior $\mathcal{R}\sim U(0,R_{\max})$ and renormalized, without rerunning the Bayesian inference.
The information gain measures the reduction in uncertainty when a broad prior is updated to a more concentrated posterior and therefore depends strongly on the adopted prior range.
In particular, \citet{buchner2022} demonstrated the relationship between the prior-to-posterior shrinkage factor and the corresponding information gain.
The results are shown in Fig.~\ref{fig:ratio_information_gain}.
For the prior $\mathcal{R} \sim U(0, 5)$, the information gain is approximately $4$ bits at early start times, decreases to a minimum at $\Delta t=8\,t_{M_z}$, where the posterior for $\mathcal{R}$ is noticeably broader, and then partially recovers.
The other prior ranges show a similar dependence on start time, with lower information gains for narrower priors.

We also examine two sources of systematic uncertainty that can affect the recovered amplitude ratio.
Figure~\ref{fig:ratio_parameter_sensitivity} illustrates the sensitivity of $\mathcal{R}$ to the adopted luminosity distance and inclination, which are varied over their median and $1\sigma$ estimates obtained from the \ac{IMR} inference of GW250114~\citep{LIGOScientific:2026ctl}.
The dependence of $\mathcal{R}$ on the analysis start time remains largely unchanged over the parameter values considered here, although its overall magnitude shifts systematically.
The observed amplitudes of the different modes scale inversely with the luminosity distance $D_L$.
Equation~\ref{eq:nonlinear_ratio} therefore implies that $\mathcal{R}$ also scales inversely with $D_L$.
Consequently, adopting a larger luminosity distance leads to a smaller inferred $\mathcal{R}$, consistent with the trend shown in the figure.
The inclination $\iota$ affects the inferred $\mathcal{R}$ through the spin-weighted spheroidal harmonics with spin weight $-2$, specifically through the magnitudes $\lvert {}_{-2}S_{22}\rvert$ and $\lvert {}_{-2}S_{44}\rvert$.
The magnitude $\lvert {}_{-2}S_{22}\rvert$ decreases with increasing $\iota$, whereas $\lvert {}_{-2}S_{44}\rvert$ initially increases and subsequently decreases, reaching its maximum at approximately $\iota=1.05$~\citep{Berti:2005gp}.
Because all the inclination values considered here are below $1.05$, the inferred $\mathcal{R}$ decreases with increasing $\iota$, again consistent with the behavior shown in the figure.
These results indicate that the present measurement should be interpreted as conditional on the adopted values of the fixed extrinsic parameters.

In addition to systematic uncertainties associated with the fixed parameters, other components of the ringdown signal may affect our analysis, including the direct waves and weaker \acp{QNM}.
The direct wave contribution is primarily present at $\Delta t\lesssim4\,t_{M_z}$ and may therefore bias analyses starting within this interval~\citep{Lu:2025vol, Oshita:2025qmn}.
Other weak \acp{QNM}, however, may affect the inference throughout the ringdown.
We investigate their impact by adding individual modes to the recovery model at $\Delta t = 5 \, t_{M_z}$, where the signal retains a relatively high \ac{SNR} and is expected to be unaffected by the direct wave.
Here, we consider the $222$ and $440$ modes as plausible additional contributions because of their relatively high expected amplitude for GW250114~\citep{Cheung:2023vki}.
The baseline three-mode model yields $\mathcal{R}=0.216^{+0.090}_{-0.092}$.
Including the $222$ mode gives $\mathcal{R}=0.223^{+0.099}_{-0.095}$, increasing the width of the central $68\%$ credible interval by approximately $6.8\%$, with little change in the posterior median.
Including the $440$ mode instead gives $\mathcal{R}=0.220^{+0.144}_{-0.130}$, with a larger interval broadening of approximately $51.0\%$.
Similar results are obtained at the other analysis start times.
The stronger impact of the $440$ mode is consistent with its oscillation frequency and damping time being closer to those of the $220 \times 220$ mode, leading to a stronger degeneracy between their contributions.

\begin{figure*}
    \centering
    \includegraphics[width=0.95\linewidth]{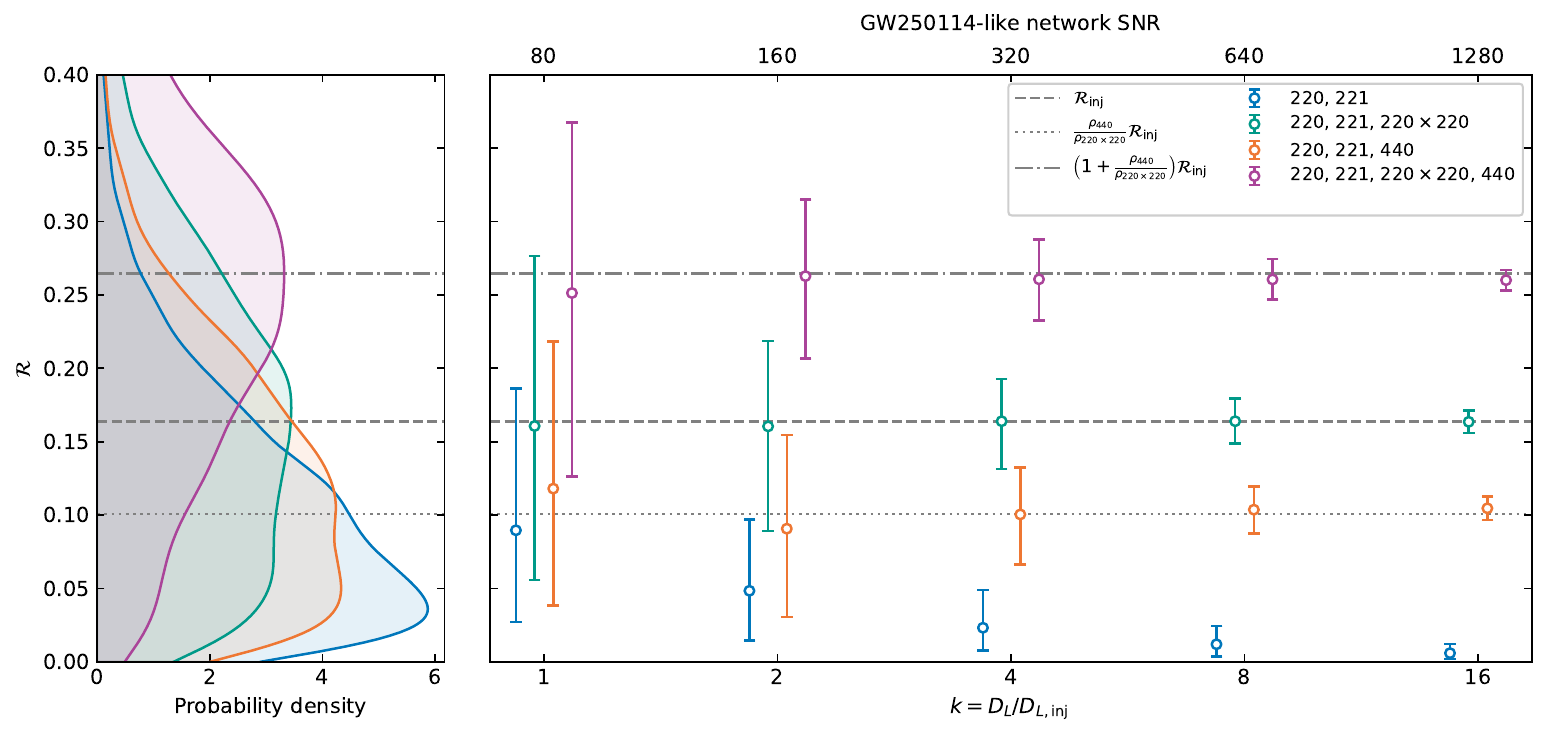}
    \caption{
        Recovery of the quadratic-mode amplitude ratio $\mathcal{R}$ as a function of signal amplitude for four zero-noise injection configurations, all analyzed with the $220,221,220\times220$ model at $\Delta t=5\,t_{M_z}$.
        Colors indicate the injected mode content.
        The left panel shows the posterior densities at the reference distance $D_L=411\,\mathrm{Mpc}$.
        The right panel shows posterior medians and central $68\%$ credible intervals as the injection distance is reduced to $D_{L,\mathrm{inj}}=D_L/k$.
        The upper axis gives the approximate equivalent full-event network SNR, scaled as $80k$ from GW250114.
        The gray dashed line marks $\mathcal{R}_{\rm inj}=0.1637$, the injected ratio for configurations containing the $220\times220$ mode.
        The dotted and dash-dotted lines show the heuristic reference values $(\rho_{440}/\rho_{220\times220})\mathcal{R}_{\rm inj}$ and $(1+\rho_{440}/\rho_{220\times220})\mathcal{R}_{\rm inj}$, respectively, where $\rho_{440}$ and $\rho_{220\times220}$ are the individual mode SNRs.
        As the \ac{SNR} increases, the credible intervals narrow, making the systematic offsets associated with the unmodeled $440$ contribution more apparent.
    }
    \label{fig:ratio_vs_amplitude_scale}
\end{figure*}

To further investigate how the $440$ mode affects the inference of $\mathcal{R}$, we construct GW250114-like synthetic signals containing different combinations of \acp{QNM}: the $220,221$ model, the $220,221,220\times220$ model, the $220,221,440$ model, and the $220,221,220\times220,440$ model.
All signals are analyzed in zero noise using the same three-mode recovery model adopted throughout this work at $\Delta t=5\,t_{M_z}$.
The left panel of Fig.~\ref{fig:ratio_vs_amplitude_scale} shows the recovery posteriors at the reference distance $D_L=411\,\mathrm{Mpc}$.
The $220, 221$ injection yields a posterior shifted toward zero while the $220, 221, 220\times 220$ injection gives a distribution consistent with $\mathcal{R}_{\rm inj}=0.1637$.
The two configurations containing the $440$ mode yield broad posteriors that overlap substantially with the latter distribution, illustrating the difficulty of distinguishing these contributions at the reference signal amplitude.
At the current \ac{SNR}, statistical uncertainties can mask the systematic bias introduced by the unmodeled $440$ mode, making it difficult to distinguish these contributions and leaving the inferred ratio sensitive to the noise realization.

Future \ac{GW} observations with higher \acp{SNR} will enable more precise measurements of the quadratic-mode amplitude ratio, making contamination from the $440$ mode potentially more important.
We simulated an increase in \ac{SNR} by reducing the luminosity distance of the injected source.
The distance scaling factor, $k=D_L/D_{L,\mathrm{inj}}$, approximately corresponds to the factor by which the \ac{SNR} increases.
The right panel of Fig.~\ref{fig:ratio_vs_amplitude_scale} shows the recovery results for different scaling factors in zero noise simulation.
As the \ac{SNR} increases, injections containing the $440$ mode yield increasingly significant offsets from their injected amplitude ratios, exceeding $2 \sigma$ for $k\geq4$ ($\mathrm{SNR} \geq 320$).
The approximately order-of-magnitude improvement in strain sensitivity targeted by third-generation ground-based detectors, such as the Einstein Telescope~\citep{ET:2019dnz} and Cosmic Explorer~\citep{Evans:2021gyd}, could yield a comparable increase in \ac{SNR} for otherwise similar sources.
These prospects highlight the need to account for other potential \ac{QNM} contributions, including the $440$ mode, when making more precise measurements of the quadratic-mode coupling coefficient in the future.

\section{Discussion}\label{sec:discussion}

In this work, we have presented the first observational test of the amplitude ratio between a \ac{QQNM} and its source \ac{QNM}.
Using the post-peak data of GW250114, we performed Bayesian inference with a ringdown model containing three modes: $220$, $221$, and $220\times220$.
We measured the amplitude ratio between the strongest \ac{QQNM}, the $220 \times 220$ mode, and its source $220$ mode, and compared it with several representative theoretical predictions.

Our results are consistent with current theoretical predictions.
The amplitude ratio distribution inferred from the observed data is broad and shows dependence on the analysis start time, reflecting that the \ac{SNR} in the post-peak data is still not sufficient for a precise measurement.
We find no significant deviation larger than $1.3 \sigma$ between the measurements and the different theoretical predictions.
At the same time, the current data do not allow us to determine which spin-dependent model best describes this event.

For GW250114, within the $44$ multipole, the \ac{SNR} of the $440$ mode is comparable to that of the $220\times220$ mode~\citep{Stein:2019qnm,London:2014cma,Cheung:2022rbm,Yang:2025ror}.
As a robustness check, we repeated the analysis using a four-mode model that additionally includes the $440$ mode.
The resulting posteriors remain consistent with those from the baseline three-mode analysis and do not materially alter our main conclusion, although the additional free parameters broaden the posterior distributions.
We therefore adopt the three-mode analysis as our fiducial result.
This choice does not resolve the ambiguity in the mode content, and the inferred coupling coefficient remains conditional on the assumed presence of the quadratic mode.
Injection tests using simulated data nevertheless indicate that the presence of the $440$ mode may bias the inferred $\mathcal{R}$ toward larger values.
This potential systematic bias remains subdominant to the statistical uncertainty and therefore does not affect our main conclusion.
However, this effect may no longer be negligible in future observations with higher \acp{SNR}, particularly at least 4 times that of GW250114 (about 320).

Our analysis still has several limitations.
First, in the Bayesian inference, we fix the luminosity distance $D_L$ and the inclination $\iota$, because they are strongly degenerate with the amplitude parameters that we aim to measure.
Tests in which these fixed values are varied show that the inferred $\mathcal{R}$ is sensitive to their choice, particularly to the adopted inclination.
Our measurement should therefore be interpreted as conditional on the fixed extrinsic parameters.
This limitation could be addressed in future work using a semi-coherent approach similar to that adopted by \citet{Wang:2026rev}, in which information from the inspiral phase is incorporated to jointly infer $D_L$, $\iota$, and the ringdown parameters, thereby enabling a more robust measurement.
Second, the present analysis is based on a single event, so the constraint on the spin dependence of the amplitude ratio is limited to a narrow range of parameter space.
In the future, especially after more high \ac{SNR} \ac{GW} events are observed, multiple events could be combined to constrain this dependence more tightly.

\begin{acknowledgments}

We thank Jiandong Zhang, Hongyu Chen, XinYi Che, Tangchao Zhan and Yupeng Wang for useful discussions on black-hole ringdown, nonlinear quasinormal modes, and gravitational-wave data analysis. 
This work has been supported in part by the National Key Research and Development Program of China (No. 2023YFC2206701) and the Fundamental Research Funds for the Central Universities, Sun Yat-sen University. 
CHS has been supported  by the Natural Science Foundation of China (Grants  No.12405080).
YMH has also been supported by the science research grants from the China Manned Space Project.
The authors acknowledge the use of ChatGPT \citep{OpenAIChatGPT} and Codex \citep{OpenAICodex}, both developed by OpenAI, for assistance with literature exploration, code development, and linguistic refinement of the manuscript. All AI-assisted outputs were critically assessed and independently verified by the authors, who take full responsibility for the scientific content and conclusions of this work.

\end{acknowledgments}

\begin{contribution}


Yuxin Yang performed the data analysis, generated the figures, and drafted the manuscript.
Chang-Fu Shi designed the analysis, validated the results, and supervised the project.
Yi-Ming Hu validated the results, supervised the project, and provided computational and funding support.
All authors discussed the results and approved the final manuscript.

\end{contribution}

%
\facilities{LIGO, Virgo, KAGRA}

\software{Bilby, dynesty, NumPy, SciPy, Matplotlib, GetDist}


\appendix

\begin{figure*}[htbp]
    \centering
    \begin{interactive}{js}{fig06int.zip}
        \includegraphics[width=0.95\linewidth]
        {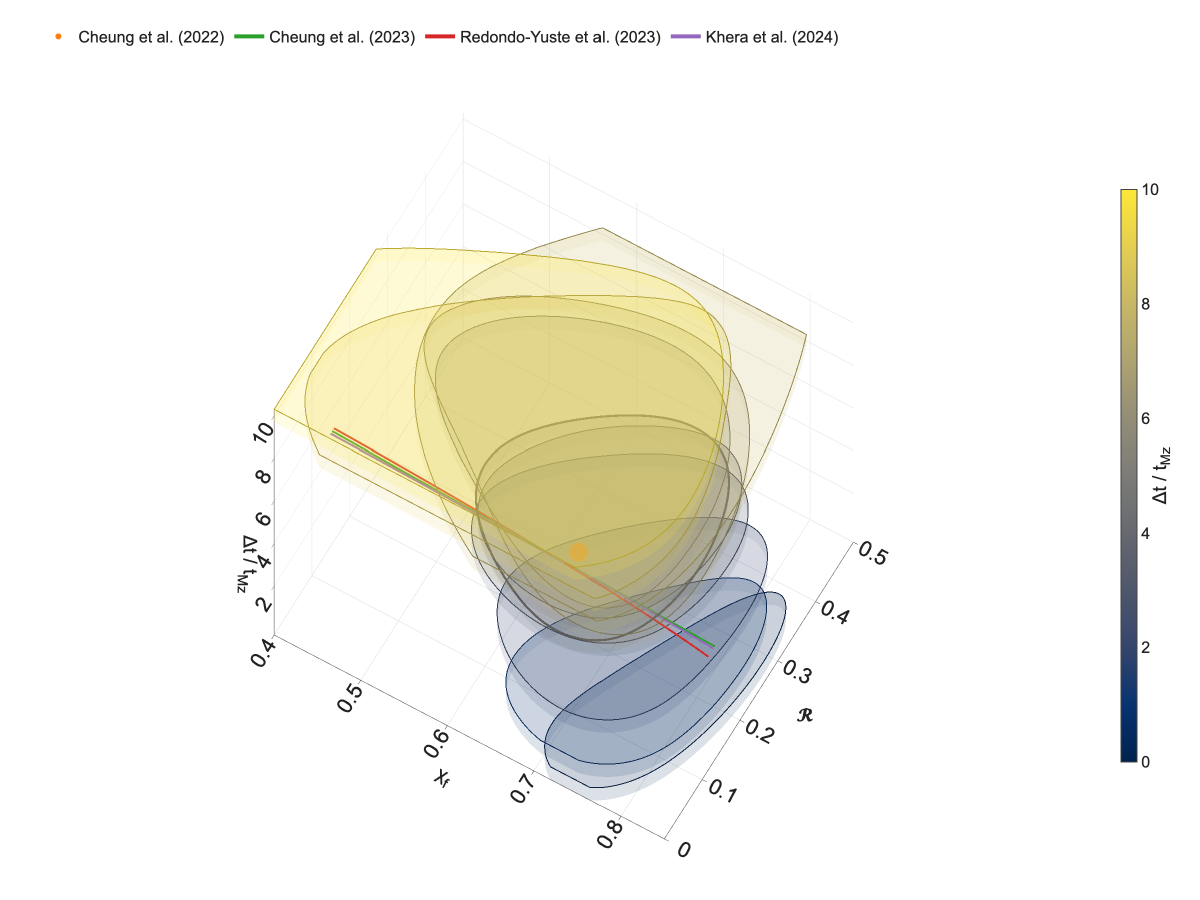}
    \end{interactive}
    \caption{
        Three-dimensional visualization of the posterior constraints shown in Fig.~\ref{fig:ratio_chi_theory_overlay}.
        The posterior regions are displayed at their respective analysis start times, with the $\Delta t=5\,t_{M_z}$ contour highlighted.
        An interactive HTML/JavaScript version is available in the online article, allowing readers to rotate and zoom the view and selectively display the posterior regions and theoretical predictions.
    }
    \label{fig:posterior_3d}
\end{figure*}

Figure~\ref{fig:posterior_3d} provides a three-dimensional representation of the posterior constraints shown in Fig.~\ref{fig:ratio_chi_theory_overlay}, with the analysis start time displayed along an additional axis.

\bibliography{sample701}{}

@article{Berti:2005gp,
    author = "Berti, Emanuele and Cardoso, Vitor and Casals, Marc",
    title = "{Eigenvalues and eigenfunctions of spin-weighted spheroidal harmonics in four and higher dimensions}",
    eprint = "gr-qc/0511111",
    archivePrefix = "arXiv",
    doi = "10.1103/PhysRevD.73.109902",
    journal = "Phys. Rev. D",
    volume = "73",
    pages = "024013",
    year = "2006",
    note = "[Erratum: Phys.Rev.D 73, 109902 (2006)]"
}

@article{Teukolsky:1973ha,
    author = "Teukolsky, Saul A.",
    title = "{Perturbations of a rotating black hole. 1. Fundamental equations for gravitational electromagnetic and neutrino field perturbations}",
    doi = "10.1086/152444",
    journal = "Astrophys. J.",
    volume = "185",
    pages = "635--647",
    year = "1973"
}

@article{Berti:2009kk,
    author = "Berti, Emanuele and Cardoso, Vitor and Starinets, Andrei O.",
    title = "{Quasinormal modes of black holes and black branes}",
    eprint = "0905.2975",
    archivePrefix = "arXiv",
    primaryClass = "gr-qc",
    doi = "10.1088/0264-9381/26/16/163001",
    journal = "Class. Quant. Grav.",
    volume = "26",
    pages = "163001",
    year = "2009"
}

@article{Mitman:2022qdl,
    author = "Mitman, Keefe and others",
    title = "{Nonlinearities in Black Hole Ringdowns}",
    eprint = "2208.07380",
    archivePrefix = "arXiv",
    primaryClass = "gr-qc",
    doi = "10.1103/PhysRevLett.130.081402",
    journal = "Phys. Rev. Lett.",
    volume = "130",
    number = "8",
    pages = "081402",
    year = "2023"
}

@article{Cheung:2022rbm,
    author = "Cheung, Mark Ho-Yeuk and others",
    title = "{Nonlinear Effects in Black Hole Ringdown}",
    eprint = "2208.07374",
    archivePrefix = "arXiv",
    primaryClass = "gr-qc",
    doi = "10.1103/PhysRevLett.130.081401",
    journal = "Phys. Rev. Lett.",
    volume = "130",
    number = "8",
    pages = "081401",
    year = "2023"
}

@article{Nakano:2007cj,
    author = "Nakano, Hiroyuki and Ioka, Kunihito",
    title = "{Second Order Quasi-Normal Mode of the Schwarzschild Black Hole}",
    eprint = "0708.0450",
    archivePrefix = "arXiv",
    primaryClass = "gr-qc",
    doi = "10.1103/PhysRevD.76.084007",
    journal = "Phys. Rev. D",
    volume = "76",
    pages = "084007",
    year = "2007"
}

@article{Cheung:2023vki,
    author = "Cheung, Mark Ho-Yeuk and Berti, Emanuele and Baibhav, Vishal and Cotesta, Roberto",
    title = "{Extracting linear and nonlinear quasinormal modes from black hole merger simulations}",
    eprint = "2310.04489",
    archivePrefix = "arXiv",
    primaryClass = "gr-qc",
    doi = "10.1103/PhysRevD.109.044069",
    journal = "Phys. Rev. D",
    volume = "109",
    number = "4",
    pages = "044069",
    year = "2024",
    note = "[Erratum: Phys.Rev.D 110, 049902 (2024), Erratum: Phys.Rev.D 112, 049901 (2025)]"
}

@article{Redondo-Yuste:2023seq,
    author = "Redondo-Yuste, Jaime and Carullo, Gregorio and Ripley, Justin L. and Berti, Emanuele and Cardoso, Vitor",
    title = "{Spin dependence of black hole ringdown nonlinearities}",
    eprint = "2308.14796",
    archivePrefix = "arXiv",
    primaryClass = "gr-qc",
    doi = "10.1103/PhysRevD.109.L101503",
    journal = "Phys. Rev. D",
    volume = "109",
    number = "10",
    pages = "L101503",
    year = "2024"
}

@article{Bucciotti:2024zyp,
    author = "Bucciotti, Bruno and Juliano, Leonardo and Kuntz, Adrien and Trincherini, Enrico",
    title = "{Quadratic quasinormal modes of a Schwarzschild black hole}",
    eprint = "2405.06012",
    archivePrefix = "arXiv",
    primaryClass = "gr-qc",
    doi = "10.1103/PhysRevD.110.104048",
    journal = "Phys. Rev. D",
    volume = "110",
    number = "10",
    pages = "104048",
    year = "2024"
}

@article{Ma:2024qcv,
    author = "Ma, Sizheng and Yang, Huan",
    title = "{Excitation of quadratic quasinormal modes for Kerr black holes}",
    eprint = "2401.15516",
    archivePrefix = "arXiv",
    primaryClass = "gr-qc",
    doi = "10.1103/PhysRevD.109.104070",
    journal = "Phys. Rev. D",
    volume = "109",
    number = "10",
    pages = "104070",
    year = "2024"
}

@article{Zhu:2024rej,
    author = "Zhu, Hengrui and others",
    title = "{Nonlinear effects in black hole ringdown from scattering experiments: Spin and initial data dependence of quadratic mode coupling}",
    eprint = "2401.00805",
    archivePrefix = "arXiv",
    primaryClass = "gr-qc",
    doi = "10.1103/PhysRevD.109.104050",
    journal = "Phys. Rev. D",
    volume = "109",
    number = "10",
    pages = "104050",
    year = "2024"
}

@article{Bucciotti:2024jrv,
    author = "Bucciotti, Bruno and Juliano, Leonardo and Kuntz, Adrien and Trincherini, Enrico",
    title = "{Amplitudes and polarizations of quadratic quasi-normal modes for a Schwarzschild black hole}",
    eprint = "2406.14611",
    archivePrefix = "arXiv",
    primaryClass = "hep-th",
    doi = "10.1007/JHEP09(2024)119",
    journal = "JHEP",
    volume = "09",
    pages = "119",
    year = "2024"
}

@article{Bourg:2024jme,
    author = "Bourg, Patrick and Panosso Macedo, Rodrigo and Spiers, Andrew and Leather, Benjamin and Bonga, B{\'e}atrice and Pound, Adam",
    title = "{Quadratic Quasinormal Mode Dependence on Linear Mode Parity}",
    eprint = "2405.10270",
    archivePrefix = "arXiv",
    primaryClass = "gr-qc",
    doi = "10.1103/PhysRevLett.134.061401",
    journal = "Phys. Rev. Lett.",
    volume = "134",
    number = "6",
    pages = "061401",
    year = "2025"
}

@article{Khera:2024bjs,
    author = "Khera, Neev and Ma, Sizheng and Yang, Huan",
    title = "{Quadratic Mode Couplings in Rotating Black Holes and Their Detectability}",
    eprint = "2410.14529",
    archivePrefix = "arXiv",
    primaryClass = "gr-qc",
    doi = "10.1103/PhysRevLett.134.211404",
    journal = "Phys. Rev. Lett.",
    volume = "134",
    number = "21",
    pages = "211404",
    year = "2025"
}

@article{Mitman:2025hgy,
    author = "Mitman, Keefe and others",
    title = "{Probing the ringdown perturbation in binary black hole coalescences with an improved quasinormal mode extraction algorithm}",
    eprint = "2503.09678",
    archivePrefix = "arXiv",
    primaryClass = "gr-qc",
    doi = "10.1103/qq1g-jlnw",
    journal = "Phys. Rev. D",
    volume = "112",
    number = "6",
    pages = "064016",
    year = "2025"
}

@unpublished{Isi:2021iql,
    author = "Isi, Maximiliano and Farr, Will M.",
    title = "{Analyzing black-hole ringdowns}",
    eprint = "2107.05609",
    archivePrefix = "arXiv",
    primaryClass = "gr-qc",
    reportNumber = "LIGO-P2100227",
    month = "7",
    year = "2021",
    note = ""
}

@article{LIGOScientific:2025wao,
    author = "Abac, A. G. and others",
    collaboration = "LIGO Scientific, Virgo, KAGRA",
    title = "{Black Hole Spectroscopy and Tests of General Relativity with GW250114}",
    eprint = "2509.08099",
    archivePrefix = "arXiv",
    primaryClass = "gr-qc",
    reportNumber = "LIGO P2500461",
    doi = "10.1103/6c61-fm1n",
    journal = "Phys. Rev. Lett.",
    volume = "136",
    number = "4",
    pages = "041403",
    year = "2026"
}

@article{LIGOScientific:2025rid,
    author = "Abac, A. G. and others",
    collaboration = "LIGO Scientific, Virgo, KAGRA",
    title = "{GW250114: Testing Hawking{\textquoteright}s Area Law and the Kerr Nature of Black Holes}",
    eprint = "2509.08054",
    archivePrefix = "arXiv",
    primaryClass = "gr-qc",
    reportNumber = "LIGO-P2500421",
    doi = "10.1103/kw5g-d732",
    journal = "Phys. Rev. Lett.",
    volume = "135",
    number = "11",
    pages = "111403",
    year = "2025"
}

@article{Speagle:2019ivv,
    author = "Speagle, Joshua S.",
    title = "{dynesty: a dynamic nested sampling package for estimating Bayesian posteriors and evidences}",
    eprint = "1904.02180",
    archivePrefix = "arXiv",
    primaryClass = "astro-ph.IM",
    doi = "10.1093/mnras/staa278",
    journal = "Mon. Not. Roy. Astron. Soc.",
    volume = "493",
    number = "3",
    pages = "3132--3158",
    year = "2020"
}

@software{sergey_koposov_2025_17268284,
  author       = {Sergey Koposov and
                  Josh Speagle and
                  Kyle Barbary and
                  Gregory Ashton and
                  Ed Bennett and
                  Johannes Buchner and
                  Carl Scheffler and
                  Colm Talbot and
                  Ben Cook and
                  James Guillochon and
                  Patricio Cubillos and
                  Andr{\'e}s Asensio Ramos and
                  Matthieu Dartiailh and
                  Ilya and
                  Erik Tollerud and
                  Dustin Lang and
                  Ben Johnson and
                  jtmendel and
                  Edward Higson and
                  Thomas Vandal and
                  Tansu Daylan and
                  Ruth Angus and
                  patelR and
                  Phillip Cargile and
                  Patrick Sheehan and
                  Matt Pitkin and
                  Matthew Kirk and
                  Lu Xu and
                  Joel Leja and
                  joezuntz},
  title        = {joshspeagle/dynesty: v3.0.0},
  month        = oct,
  year         = 2025,
  publisher    = {Zenodo},
  version      = {v3.0.0},
  doi          = {10.5281/zenodo.17268284},
  url          = {https://doi.org/10.5281/zenodo.17268284},
  swhid        = {swh:1:dir:b3340f6aaa931bae6c6bcff6e7ddeb89ca508a15
                   ;origin=https://doi.org/10.5281/zenodo.3348367;vis
                   it=swh:1:snp:fa92422c27d5611b107216e9b766a871bb43b
                   bee;anchor=swh:1:rel:94af3a8ef6e50d997258e152c3833
                   3550bd7463f;path=joshspeagle-dynesty-217bc94
                  },
}

@software{Bilby_17371955,
  author       = {Colm Talbot and
                  Gregory Ashton and
                  Moritz H{\"u}bner and
                  Matt Pitkin and
                  plasky and
                  asb5468 and
                  Michael J. Williams and
                  Aditya Vijaykumar and
                  Rory Smith and
                  SMorisaki and
                  John Veitch and
                  Nikhil Sarin and
                  Duncan Macleod and
                  Daniel Williams and
                  JasperMartins and
                  MarcArene and
                  C P L Berry and
                  Vivien Raymond and
                  Ceciliogq and
                  Ivan Markin and
                  David Keitel and
                  AlexandreGoettel and
                  Lorenzo Pompili and
                  Mick Wright and
                  oliviawilk and
                  noahewolfe and
                  jacobgolomb and
                  Shichao Wu and
                  Rhiannon Udall and
                  Michael P{\"u}rrer},
  title        = {bilby-dev/bilby: v2.7.0},
  month        = oct,
  year         = 2025,
  publisher    = {Zenodo},
  version      = {v2.7.0},
  doi          = {10.5281/zenodo.17371955},
  url          = {https://doi.org/10.5281/zenodo.17371955},
  swhid        = {swh:1:dir:b5e7d3fdbeef9917c38d5574f69784bcea5e4d9d
                   ;origin=https://doi.org/10.5281/zenodo.14025463;vi
                   sit=swh:1:snp:5a2654492f8e8853fbb6871580a33ae3ee96
                   249e;anchor=swh:1:rel:43931a413f362fe16249656762c1
                   2266f5754c59;path=bilby-dev-bilby-8e52c03
                  },
}

@article{bilby_paper,
       author = {{Ashton}, Gregory and {H{\"u}bner}, Moritz and {Lasky}, Paul D. and {Talbot}, Colm and {Ackley}, Kendall and {Biscoveanu}, Sylvia and {Chu}, Qi and {Divakarla}, Atul and {Easter}, Paul J. and {Goncharov}, Boris and {Hernandez Vivanco}, Francisco and {Harms}, Jan and {Lower}, Marcus E. and {Meadors}, Grant D. and {Melchor}, Denyz and {Payne}, Ethan and {Pitkin}, Matthew D. and {Powell}, Jade and {Sarin}, Nikhil and {Smith}, Rory J.~E. and {Thrane}, Eric},
        title = "{BILBY: A User-friendly Bayesian Inference Library for Gravitational-wave Astronomy}",
      journal = {Astrophys. J. Suppl.},
         year = 2019,
        month = apr,
       volume = {241},
       number = {2},
          eid = {27},
        pages = {27},
          doi = {10.3847/1538-4365/ab06fc},
archivePrefix = {arXiv},
       eprint = {1811.02042},
 primaryClass = {astro-ph.IM},
       adsurl = {https://ui.adsabs.harvard.edu/abs/2019ApJS..241...27A}
}

@article{LIGOScientific:2016aoc,
    author = "Abbott, B. P. and others",
    collaboration = "LIGO Scientific, Virgo",
    title = "{Observation of Gravitational Waves from a Binary Black Hole Merger}",
    eprint = "1602.03837",
    archivePrefix = "arXiv",
    primaryClass = "gr-qc",
    reportNumber = "LIGO-P150914",
    doi = "10.1103/PhysRevLett.116.061102",
    journal = "Phys. Rev. Lett.",
    volume = "116",
    number = "6",
    pages = "061102",
    year = "2016"
}

@article{LIGOScientific:2018mvr,
    author = "Abbott, B. P. and others",
    collaboration = "LIGO Scientific, Virgo",
    title = "{GWTC-1: A Gravitational-Wave Transient Catalog of Compact Binary Mergers Observed by LIGO and Virgo during the First and Second Observing Runs}",
    eprint = "1811.12907",
    archivePrefix = "arXiv",
    primaryClass = "astro-ph.HE",
    reportNumber = "LIGO-P1800307",
    doi = "10.1103/PhysRevX.9.031040",
    journal = "Phys. Rev. X",
    volume = "9",
    number = "3",
    pages = "031040",
    year = "2019"
}

@article{LIGOScientific:2020ibl,
    author = "Abbott, R. and others",
    collaboration = "LIGO Scientific, Virgo",
    title = "{GWTC-2: Compact Binary Coalescences Observed by LIGO and Virgo During the First Half of the Third Observing Run}",
    eprint = "2010.14527",
    archivePrefix = "arXiv",
    primaryClass = "gr-qc",
    reportNumber = "P2000061",
    doi = "10.1103/PhysRevX.11.021053",
    journal = "Phys. Rev. X",
    volume = "11",
    pages = "021053",
    year = "2021"
}

@article{KAGRA:2021vkt,
    author = "Abbott, R. and others",
    collaboration = "KAGRA, VIRGO, LIGO Scientific",
    title = "{GWTC-3: Compact Binary Coalescences Observed by LIGO and Virgo during the Second Part of the Third Observing Run}",
    eprint = "2111.03606",
    archivePrefix = "arXiv",
    primaryClass = "gr-qc",
    reportNumber = "LIGO-P2000318",
    doi = "10.1103/PhysRevX.13.041039",
    journal = "Phys. Rev. X",
    volume = "13",
    number = "4",
    pages = "041039",
    year = "2023"
}

@unpublished{LIGOScientific:2025slb,
    author = "Abac, A. G. and others",
    collaboration = "LIGO Scientific, VIRGO, KAGRA",
    title = "{GWTC-4.0: Updating the Gravitational-Wave Transient Catalog with Observations from the First Part of the Fourth LIGO-Virgo-KAGRA Observing Run}",
    eprint = "2508.18082",
    archivePrefix = "arXiv",
    primaryClass = "gr-qc",
    reportNumber = "LIGO-P2400386",
    month = "8",
    year = "2025",
    note = ""
}

@article{Kokkotas:1999bd,
    author = "Kokkotas, Kostas D. and Schmidt, Bernd G.",
    title = "{Quasinormal modes of stars and black holes}",
    eprint = "gr-qc/9909058",
    archivePrefix = "arXiv",
    doi = "10.12942/lrr-1999-2",
    journal = "Living Rev. Rel.",
    volume = "2",
    pages = "2",
    year = "1999"
}

@article{Nollert:1999ji,
    author = "Nollert, Hans-Peter",
    title = "{TOPICAL REVIEW: Quasinormal modes: the characteristic `sound' of black holes and neutron stars}",
    doi = "10.1088/0264-9381/16/12/201",
    journal = "Class. Quant. Grav.",
    volume = "16",
    pages = "R159--R216",
    year = "1999"
}

@article{Berti:2005ys,
    author = "Berti, Emanuele and Cardoso, Vitor and Will, Clifford M.",
    title = "{On gravitational-wave spectroscopy of massive black holes with the space interferometer LISA}",
    eprint = "gr-qc/0512160",
    archivePrefix = "arXiv",
    doi = "10.1103/PhysRevD.73.064030",
    journal = "Phys. Rev. D",
    volume = "73",
    pages = "064030",
    year = "2006"
}

@article{Gossan:2011ha,
    author = "Gossan, S. and Veitch, J. and Sathyaprakash, B. S.",
    title = "{Bayesian model selection for testing the no-hair theorem with black hole ringdowns}",
    eprint = "1111.5819",
    archivePrefix = "arXiv",
    primaryClass = "gr-qc",
    doi = "10.1103/PhysRevD.85.124056",
    journal = "Phys. Rev. D",
    volume = "85",
    pages = "124056",
    year = "2012"
}

@article{Dreyer2004,
author = {Dreyer, Olaf and Kelly, Bernard and Krishnan, Badri and Finn, Lee Samuel and Garrison, David and Lopez-Aleman, Ramon},
doi = {10.1088/0264-9381/21/4/003},
eprint = {0309007},
issn = {02649381},
journal = {Classical and Quantum Gravity},
number = {4},
pages = {787--803},
primaryClass = {gr-qc},
title = {{Black-hole spectroscopy: Testing general relativity through gravitational-wave observations}},
volume = {21},
year = {2004}
}

@article{Shi:2019hqa,
    author = "Shi, Changfu and Bao, Jiahui and Wang, Haitian and Zhang, Jian-dong and Hu, Yiming and Sesana, Alberto and Barausse, Enrico and Mei, Jianwei and Luo, Jun",
    title = "{Science with the TianQin observatory: Preliminary results on testing the no-hair theorem with ringdown signals}",
    eprint = "1902.08922",
    archivePrefix = "arXiv",
    primaryClass = "gr-qc",
    doi = "10.1103/PhysRevD.100.044036",
    journal = "Phys. Rev. D",
    volume = "100",
    number = "4",
    pages = "044036",
    year = "2019"
}

@unpublished{LIGOScientific:2026wpt,
    author = "Abac, A. G. and others",
    collaboration = "LIGO Scientific, VIRGO, KAGRA",
    title = "{GWTC-4.0: Tests of General Relativity. III. Tests of the Remnants}",
    eprint = "2603.19021",
    archivePrefix = "arXiv",
    primaryClass = "gr-qc",
    reportNumber = "LIGO-P2500067",
    month = "3",
    year = "2026",
    note = ""
}

@article{London:2014cma,
    author = "London, Lionel and Shoemaker, Deirdre and Healy, James",
    title = "{Modeling ringdown: Beyond the fundamental quasinormal modes}",
    eprint = "1404.3197",
    archivePrefix = "arXiv",
    primaryClass = "gr-qc",
    doi = "10.1103/PhysRevD.90.124032",
    journal = "Phys. Rev. D",
    volume = "90",
    number = "12",
    pages = "124032",
    year = "2014",
    note = "[Erratum: Phys.Rev.D 94, 069902 (2016)]"
}

@unpublished{Yang:2025ror,
    author = "Yang, Yuxin and Shi, Changfu and Hu, Yi-Ming",
    title = "{Contribution from Nonlinear Quasi-normal Modes in GW250114}",
    eprint = "2510.16903",
    archivePrefix = "arXiv",
    primaryClass = "gr-qc",
    month = "10",
    year = "2025",
    note = ""
}

@unpublished{Wang:2026rev,
    author = "Wang, Yi-Fan and Ma, Sizheng and Khera, Neev and Yang, Huan",
    title = "{A nonlinear voice from GW250114 ringdown}",
    eprint = "2601.05734",
    archivePrefix = "arXiv",
    primaryClass = "gr-qc",
    reportNumber = "LIGO-P2500804",
    month = "1",
    year = "2026",
    note = ""
}

@article{Capano:2021etf,
    author = "Capano, Collin D. and Cabero, Miriam and Westerweck, Julian and Abedi, Jahed and Kastha, Shilpa and Nitz, Alexander H. and Wang, Yi-Fan and Nielsen, Alex B. and Krishnan, Badri",
    title = "{Multimode Quasinormal Spectrum from a Perturbed Black Hole}",
    eprint = "2105.05238",
    archivePrefix = "arXiv",
    primaryClass = "gr-qc",
    doi = "10.1103/PhysRevLett.131.221402",
    journal = "Phys. Rev. Lett.",
    volume = "131",
    number = "22",
    pages = "221402",
    year = "2023"
}

@article{LIGOScientific:2020iuh,
    author = "Abbott, R. and others",
    collaboration = "LIGO Scientific, Virgo",
    title = "{GW190521: A Binary Black Hole Merger with a Total Mass of $150  M_{\odot}$}",
    eprint = "2009.01075",
    archivePrefix = "arXiv",
    primaryClass = "gr-qc",
    doi = "10.1103/PhysRevLett.125.101102",
    journal = "Phys. Rev. Lett.",
    volume = "125",
    number = "10",
    pages = "101102",
    year = "2020"
}

@article{Siegel:2023lxl,
    author = "Siegel, Harrison and Isi, Maximiliano and Farr, Will M.",
    title = "{Ringdown of GW190521: Hints of multiple quasinormal modes with a precessional interpretation}",
    eprint = "2307.11975",
    archivePrefix = "arXiv",
    primaryClass = "gr-qc",
    reportNumber = "LIGO-P2300214",
    doi = "10.1103/PhysRevD.108.064008",
    journal = "Phys. Rev. D",
    volume = "108",
    number = "6",
    pages = "064008",
    year = "2023"
}

@article{Wang:2025baj,
    author = "Wang, Hai-Tian",
    title = "{Decisive Evidence for the First Overtone Mode in the Ringdown Signal of GW231028}",
    eprint = "2509.08657",
    archivePrefix = "arXiv",
    primaryClass = "gr-qc",
    month = "9",
    year = "2025",
    journal = ""
}

@article{Wang:2025rvn,
    author = "Wang, Hai-Tian and Tang, Shao-Peng and Li, Peng-Cheng and Fan, Yi-Zhong",
    title = "{Detection of a Higher Harmonic Quasi-normal Mode in the Ringdown Signal of GW231123}",
    eprint = "2509.02047",
    archivePrefix = "arXiv",
    primaryClass = "gr-qc",
    month = "9",
    year = "2025",
    journal = ""
}

@article{Tang:2025jyj,
    author = "Tang, Shao-Peng and Wang, Hai-Tian and Li, Yin-Jie and Fan, Yi-Zhong",
    title = "{Verification of the black hole area law with GW230814}",
    eprint = "2509.03480",
    archivePrefix = "arXiv",
    primaryClass = "gr-qc",
    doi = "10.1016/j.scib.2025.11.002",
    journal = "Sci. Bull.",
    volume = "71",
    pages = "83--88",
    year = "2026"
}

@article{Isi:2019aib,
    author = "Isi, Maximiliano and Giesler, Matthew and Farr, Will M. and Scheel, Mark A. and Teukolsky, Saul A.",
    title = "{Testing the no-hair theorem with GW150914}",
    eprint = "1905.00869",
    archivePrefix = "arXiv",
    primaryClass = "gr-qc",
    reportNumber = "LIGO-P1900135",
    doi = "10.1103/PhysRevLett.123.111102",
    journal = "Phys. Rev. Lett.",
    volume = "123",
    number = "11",
    pages = "111102",
    year = "2019"
}

@article{Cotesta:2022pci,
    author = "Cotesta, Roberto and Carullo, Gregorio and Berti, Emanuele and Cardoso, Vitor",
    title = "{Analysis of Ringdown Overtones in GW150914}",
    eprint = "2201.00822",
    archivePrefix = "arXiv",
    primaryClass = "gr-qc",
    doi = "10.1103/PhysRevLett.129.111102",
    journal = "Phys. Rev. Lett.",
    volume = "129",
    number = "11",
    pages = "111102",
    year = "2022"
}

@unpublished{Isi:2022mhy,
    author = "Isi, Maximiliano and Farr, Will M.",
    title = "{Revisiting the ringdown of GW150914}",
    eprint = "2202.02941",
    archivePrefix = "arXiv",
    primaryClass = "gr-qc",
    reportNumber = "LIGO-P2200028",
    month = "2",
    year = "2022",
    note = ""
}

@article{Wang:2024yhb,
    author = "Wang, Hai-Tian and Wang, Ziming and Dong, Yiming and Yim, Garvin and Shao, Lijing",
    title = "{Reanalyzing the ringdown signal of GW150914 using the F-statistic method}",
    eprint = "2411.13333",
    archivePrefix = "arXiv",
    primaryClass = "gr-qc",
    doi = "10.1103/PhysRevD.111.064037",
    journal = "Phys. Rev. D",
    volume = "111",
    number = "6",
    pages = "064037",
    year = "2025"
}

@article{Wang:2023ljx,
    author = "Wang, Yi-Fan and Capano, Collin D. and Abedi, Jahed and Kastha, Shilpa and Krishnan, Badri and Nielsen, Alex B. and Nitz, Alexander H. and Westerweck, Julian",
    title = "{Gating-and-inpainting perspective on GW150914 ringdown overtone: Understanding the data analysis systematics}",
    eprint = "2310.19645",
    archivePrefix = "arXiv",
    primaryClass = "gr-qc",
    reportNumber = "LIGO-P2300340",
    doi = "10.1103/3gqn-297f",
    journal = "Phys. Rev. D",
    volume = "112",
    number = "8",
    pages = "083023",
    year = "2025"
}

@article{Stein:2019qnm,  
 title={qnm: A Python package for calculating Kerr quasinormal modes, separation constants, and spherical-spheroidal mixing coefficients}, 
 url={https://doi.org/10.21105/joss.01683}, 
 DOI={10.21105/joss.01683}, 
 journal={Journal of Open Source Software}, 
 author={Stein, Leo}, 
 year={2019}, 
 month={Oct}, 
 pages={1683}
}

@misc{buchner2022,
      title={An intuition for physicists: information gain from experiments}, 
      author={Johannes Buchner},
      year={2022},
      eprint={2205.00009},
      archivePrefix={arXiv},
      primaryClass={cond-mat.stat-mech},
      url={https://arxiv.org/abs/2205.00009}, 
}

@article{KullbackLeibler1951,
    author  = {Kullback, Solomon and Leibler, Richard A.},
    title   = {On Information and Sufficiency},
    journal = {The Annals of Mathematical Statistics},
    volume  = {22},
    number  = {1},
    pages   = {79--86},
    year    = {1951},
    doi     = {10.1214/aoms/1177729694}
}

@unpublished{LIGOScientific:2026ctl,
    author = "Abac, A. G. and others",
    collaboration = "LIGO Scientific, VIRGO, KAGRA",
    title = "{GWTC-5.0: Population Properties of Merging Compact Binaries}",
    eprint = "2605.27226",
    archivePrefix = "arXiv",
    primaryClass = "astro-ph.HE",
    reportNumber = "LIGO-P2600045",
    month = "5",
    year = "2026",
    note = ""
}

@article{Okuzumi:2008ej,
    author = "Okuzumi, Satoshi and Ioka, Kunihito and Sakagami, Masa-aki",
    title = "{Possible Discovery of Nonlinear Tail and Quasinormal Modes in Black Hole Ringdown}",
    eprint = "0803.0501",
    archivePrefix = "arXiv",
    primaryClass = "gr-qc",
    doi = "10.1103/PhysRevD.77.124018",
    journal = "Phys. Rev. D",
    volume = "77",
    pages = "124018",
    year = "2008"
}

@article{Besson:2024adi,
    author = "Besson, J{\'e}r{\'e}my and Jaramillo, Jos{\'e} Luis",
    title = "{Quasi-normal mode expansions of black hole perturbations: a hyperboloidal Keldysh{\textquoteright}s approach}",
    eprint = "2412.02793",
    archivePrefix = "arXiv",
    primaryClass = "gr-qc",
    doi = "10.1007/s10714-025-03438-6",
    journal = "Gen. Rel. Grav.",
    volume = "57",
    number = "7",
    pages = "110",
    year = "2025"
}

@article{Silva:2024ffz,
    author = "Silva, Hector O. and Tambalo, Giovanni and Glampedakis, Kostas and Yagi, Kent and Steinhoff, Jan",
    title = "{Quasinormal modes and their excitation beyond general relativity}",
    eprint = "2404.11110",
    archivePrefix = "arXiv",
    primaryClass = "gr-qc",
    doi = "10.1103/PhysRevD.110.024042",
    journal = "Phys. Rev. D",
    volume = "110",
    number = "2",
    pages = "024042",
    year = "2024"
}

@misc{OpenAIChatGPT,
  author       = {{OpenAI}},
  title        = {ChatGPT},
  year         = {2026},
  howpublished = {\url{https://chatgpt.com/}},
  note         = {Large language model-based AI assistant, accessed 2026}
}

@misc{OpenAICodex,
  author       = {{OpenAI}},
  title        = {Codex},
  year         = {2026},
  howpublished = {\url{https://openai.com/codex/}},
  note         = {AI coding agent, accessed 2026}
}

@article{Lu:2025vol,
    author = "Lu, Neil and Ma, Sizheng and Piccinni, Ornella J. and Chen, Yanbei and Sun, Ling",
    title = "{GW250114 reveals signatures of post-merger black-hole horizon}",
    eprint = "2510.01001",
    archivePrefix = "arXiv",
    primaryClass = "gr-qc",
    doi = "10.1038/s41586-026-10696-0",
    journal = "Nature",
    volume = "655",
    number = "8121",
    pages = "71--75",
    year = "2026"
}

@article{Oshita:2025qmn,
    author = "Oshita, Naritaka and Ma, Sizheng and Chen, Yanbei and Yang, Huan",
    title = "{Probing Direct Waves in Black Hole Ringdowns}",
    eprint = "2509.09165",
    archivePrefix = "arXiv",
    primaryClass = "gr-qc",
    reportNumber = "YITP-25-141, RIKEN-iTHEMS-Report-25",
    month = "9",
    year = "2025"
}

@article{LIGOScientific:2026jgl,
    author = "Abac, A. G. and others",
    collaboration = "LIGO Scientific, VIRGO, KAGRA",
    title = "{Open Data from LIGO, Virgo, and KAGRA through the Second Part of the Fourth Observing Run}",
    eprint = "2605.27090",
    archivePrefix = "arXiv",
    primaryClass = "gr-qc",
    reportNumber = "LIGO-P2600085",
    month = "5",
    year = "2026"
}

@article{Romero-Shaw:2020owr,
    author = "Romero-Shaw, I. M. and others",
    title = "{Bayesian inference for compact binary coalescences with bilby: validation and application to the first LIGO{\textendash}Virgo gravitational-wave transient catalogue}",
    eprint = "2006.00714",
    archivePrefix = "arXiv",
    primaryClass = "astro-ph.IM",
    doi = "10.1093/mnras/staa2850",
    journal = "Mon. Not. Roy. Astron. Soc.",
    volume = "499",
    number = "3",
    pages = "3295--3319",
    year = "2020"
}

@article{Bhagwat:2017tkm,
    author = "Bhagwat, Swetha and Okounkova, Maria and Ballmer, Stefan W. and Brown, Duncan A. and Giesler, Matthew and Scheel, Mark A. and Teukolsky, Saul A.",
    title = "{On choosing the start time of binary black hole ringdowns}",
    eprint = "1711.00926",
    archivePrefix = "arXiv",
    primaryClass = "gr-qc",
    doi = "10.1103/PhysRevD.97.104065",
    journal = "Phys. Rev. D",
    volume = "97",
    number = "10",
    pages = "104065",
    year = "2018"
}

@article{Clarke:2024lwi,
    author = "Clarke, Teagan A. and others",
    title = "{Toward a self-consistent framework for measuring black hole ringdowns}",
    eprint = "2402.02819",
    archivePrefix = "arXiv",
    primaryClass = "gr-qc",
    doi = "10.1103/PhysRevD.109.124030",
    journal = "Phys. Rev. D",
    volume = "109",
    number = "12",
    pages = "124030",
    year = "2024"
}

@article{Yi:2024elj,
    author = "Yi, Sophia and Kuntz, Adrien and Barausse, Enrico and Berti, Emanuele and Cheung, Mark Ho-Yeuk and Kritos, Konstantinos and Maselli, Andrea",
    title = "{Nonlinear quasinormal mode detectability with next-generation gravitational wave detectors}",
    eprint = "2403.09767",
    archivePrefix = "arXiv",
    primaryClass = "gr-qc",
    reportNumber = "ET-0081A-24",
    doi = "10.1103/PhysRevD.109.124029",
    journal = "Phys. Rev. D",
    volume = "109",
    number = "12",
    pages = "124029",
    year = "2024"
}

@article{Baibhav:2023clw,
    author = "Baibhav, Vishal and Cheung, Mark Ho-Yeuk and Berti, Emanuele and Cardoso, Vitor and Carullo, Gregorio and Cotesta, Roberto and Del Pozzo, Walter and Duque, Francisco",
    title = "{Agnostic black hole spectroscopy: Quasinormal mode content of numerical relativity waveforms and limits of validity of linear perturbation theory}",
    eprint = "2302.03050",
    archivePrefix = "arXiv",
    primaryClass = "gr-qc",
    doi = "10.1103/PhysRevD.108.104020",
    journal = "Phys. Rev. D",
    volume = "108",
    number = "10",
    pages = "104020",
    year = "2023"
}

@article{Lagos:2024ekd,
    author = "Lagos, Macarena and Andrade, Tom{\'a}s and Rafecas-Ventosa, Jordi and Hui, Lam",
    title = "{Black hole spectroscopy with nonlinear quasinormal modes}",
    eprint = "2411.02264",
    archivePrefix = "arXiv",
    primaryClass = "gr-qc",
    doi = "10.1103/PhysRevD.111.024018",
    journal = "Phys. Rev. D",
    volume = "111",
    number = "2",
    pages = "024018",
    year = "2025"
}

@article{Konoplya:2011qq,
    author = "Konoplya, R. A. and Zhidenko, A.",
    title = "{Quasinormal modes of black holes: From astrophysics to string theory}",
    eprint = "1102.4014",
    archivePrefix = "arXiv",
    primaryClass = "gr-qc",
    doi = "10.1103/RevModPhys.83.793",
    journal = "Rev. Mod. Phys.",
    volume = "83",
    pages = "793--836",
    year = "2011"
}

@article{ET:2019dnz,
    author = "Maggiore, Michele and others",
    collaboration = "ET",
    title = "{Science Case for the Einstein Telescope}",
    eprint = "1912.02622",
    archivePrefix = "arXiv",
    primaryClass = "astro-ph.CO",
    doi = "10.1088/1475-7516/2020/03/050",
    journal = "JCAP",
    volume = "03",
    pages = "050",
    year = "2020"
}

@article{Evans:2021gyd,
    author = "Evans, Matthew and others",
    title = "{A Horizon Study for Cosmic Explorer: Science, Observatories, and Community}",
    eprint = "2109.09882",
    archivePrefix = "arXiv",
    primaryClass = "astro-ph.IM",
    reportNumber = "CE-P2100003-v7, Cosmic Explorer technical report CE-P2100003-v6",
    month = "9",
    year = "2021"
}

@article{Lewis:2019xzd,
    author = "Lewis, Antony",
    title = "{GetDist: a Python package for analysing Monte Carlo samples}",
    eprint = "1910.13970",
    archivePrefix = "arXiv",
    primaryClass = "astro-ph.IM",
    doi = "10.1088/1475-7516/2025/08/025",
    journal = "JCAP",
    volume = "08",
    pages = "025",
    year = "2025"
}

@article{Nollert:1996rf,
    author = "Nollert, Hans-Peter",
    title = "{About the significance of quasinormal modes of black holes}",
    eprint = "gr-qc/9602032",
    archivePrefix = "arXiv",
    doi = "10.1103/PhysRevD.53.4397",
    journal = "Phys. Rev. D",
    volume = "53",
    pages = "4397--4402",
    year = "1996"
}
\bibliographystyle{aasjournalv7}



\end{CJK*}
\end{document}